\newcommand{\etal}{{\it et al.}}

\documentclass{ws-ijmpa}
\usepackage{wrapfig, afterpage,epsfig}

\begin{document}

\markboth{N. Menaa} {Measuring ${\cal{B}}(D^+\to\mu^+\nu)$ and the
Pseudoscalar Decay Constant $f_{D^+}$}

%
\catchline{}{}{}{}{}
%

\title{\LARGE Measuring ${\cal{B}}(D^+\to\mu^+\nu)$ and the
Pseudoscalar Decay Constant $f_{D^+}$}

\author{\footnotesize N. Menaa \\ Representing the CLEO Collaboration}

\address{201 Physics Building, Syracuse University, Syracuse, New York 13244\\
E-mail: nmenaa@phy.syr.edu}

\maketitle


\begin{abstract}
 In 60 pb$^{-1}$ of data taken on the $\psi(3770)$ resonance with
the CLEO-c detector, we find 8 $D^+\to\mu^+\nu$ event candidates
that are mostly signal, containing only 1 estimated background.
Using this statistically compelling sample, we measure the value
of ${\cal{B}}(D^+\to\mu^+\nu)=(3.5\pm 1.4 \pm 0.6)\times 10^{-4}$,
and determine $f_{D^+}=(198\pm 41\pm 17)$ MeV.
\end{abstract}

\section{Introduction}
Measuring purely leptonic decays of heavy mesons allows the
determination of  meson decay constants, which connect measured
quantities, such as the $B\bar{B}$  mixing ratio, to CKM matrix
elements. Currently, it is not possible to determine $f_B$
experimentally from leptonic $B$ decays, so theoretical
calculations of $f_B$ must be used.

Measurements of pseudoscalar decay constants such as $f_{D^+}$
provide checks on these calculations and help discriminate among
different models.


The decay rate is given by\cite{Formula1} $ \Gamma(D^+\to l^+\nu)
= {{G_F^2}\over 8\pi}f_{D^+}^2m_l^2M_{D^+} \left(1-{m_l^2\over
M_{D^+}^2}\right)^2 \left|V_{cd}\right|^2$, where $M_{D^+}$ is the
$D^+$ mass, $m_l$ is the mass of the final state lepton, $V_{cd}$
is a CKM matrix element equal to 0.224,\cite{PDG} and $G_F$ is the
Fermi coupling constant.

\section{Analysis Technique and Event Selection }

In this study we use 60 pb$^{-1}$ of CLEO-c data recorded at the
$\psi''$ resonance (3.770 GeV)  with the CLEO-c
detector\cite{detector} at the Cornell Electron Storage Ring
(CESR). We fully reconstruct one charged $D$ meson of the produced
$D^+ D^-$ pairs as a tag. Tagging modes are fully reconstructed by
first evaluating the difference in the energy of the decay
products with the beam energy $\Delta E$. We then require the
absolute value of this difference to be within 0.02 GeV of zero,
approximately twice the r.m.s. width, and then look at the
reconstructed $D^-$ beam constrained mass defined as
$m_D=\sqrt{E_{beam}^2-(\sum_i\overrightarrow{p}_{\!i})^2}$, where
$i$ runs over all the final state particles. We then search for
$D^+\to \mu^+ \nu_{\mu}$ candidates with charge opposite to the
charge of the tagged $D^-$. The charge-conjugate modes are always
used.
\begin{figure}[hbtp]
\centerline{\psfig{file=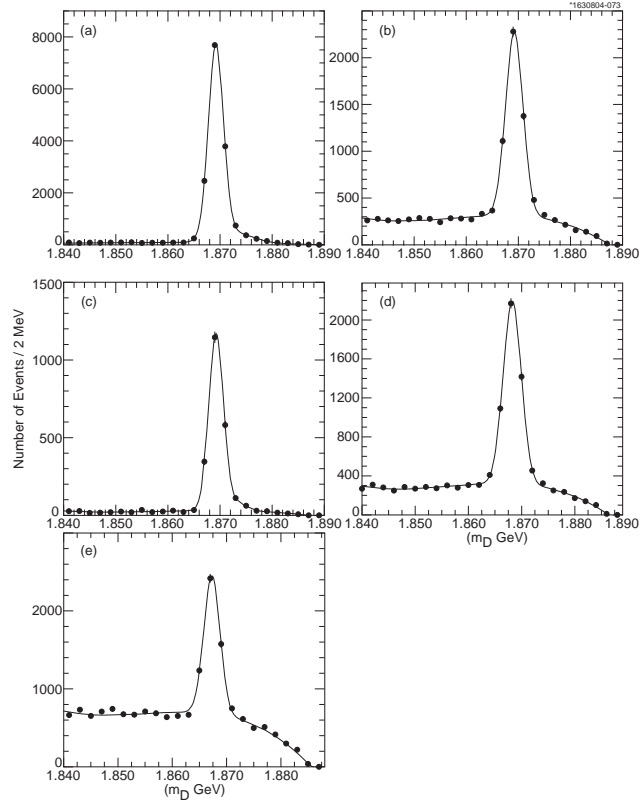,width=0.7\textwidth}}
\vspace*{8pt} \caption{Beam Constrained Mass distributions for
different fully reconstructed $D^-$ decay candidates; the curves
show the sum of Gaussian signal functions and $3^{rd}$ order
polynomial background functions. A single signal Gaussian is used
for all modes except for (a) and (c) where two Gaussians are used.
(a) $D^- \to K^+ \pi^- \pi^-$, (b) $D^-\to K^+ \pi^- \pi^- \pi^0$,
(c) $D^- \to K_s \pi^-$, (d) $D^- \to K_s \pi^-\pi^-\pi^+$ and (e)
$D^- \to K_s\pi^- \pi^0$.} \label{Drecon}
\end{figure}

The $m_D$ distributions for all $D^-$ tagging modes considered in
this data sample are shown in Fig.~\ref{Drecon}. We have a total
of 28574$\pm$207 tags.

To select $D^+\to \mu^+\nu_{\mu}$ events we first reconstruct
$D^-$ event candidates and then search for events with a single
additional charged track presumed to be a $\mu^+$. Then  we infer
the existence of the neutrino by requiring a measured value near
zero (the neutrino mass) of the missing mass squared (MM$^2$)
defined as ${\rm
MM}^2=\left(E_{beam}-E_{\mu^+}\right)^2-\left(-\overrightarrow{p}_{\!D^-}
-\overrightarrow{p}_{\!\mu^+}\right)^2 $  where
$\overrightarrow{p}_{D^-}$ is the three-momentum of the fully
reconstructed $D^-$. We reject events with extra charged tracks
besides the muon candidate and maximum extra shower energy above
250 MeV.
\section{Results}
\begin{figure}
\centerline{\psfig{file=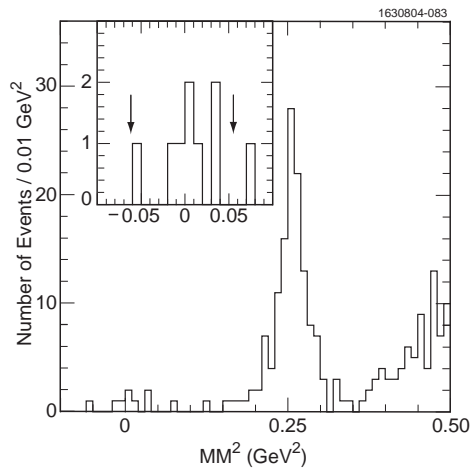,width=0.5\textwidth}}
\caption{MM$^2$ using $D^-$ tags and one additional opposite sign
charged track and no extra energetic showers (see text). The
insert shows the signal region for $D^+\to\mu^+\nu$ enlarged, $\pm
2\sigma$ range is shown between the two arrows.} \label{mm2}
\end{figure}

The MM$^2$ distribution for our tagged events is shown in
Fig.~\ref{mm2}.\cite{ICHEP:Nabil} We see a signal near zero
containing 8 events within a $\pm2\sigma$ interval, -0.056 GeV$^2$
to +0.056 GeV$^2$. This signal is due to the
$D^+\to\mu^+\nu_{\mu}$ mode we are seeking. The large peak
centered near 0.25 GeV$^2$ is from the decay $D^+\to
\overline{K}^o\pi^+$ that is far from our signal region.

Backgrounds from $D^+\to\tau^+\nu$, $\tau^+\to \pi^+\nu$,
$D^+\to\pi^+\pi^0$ and other smaller sources are estimated as
summing to 1.07 events.
 Subtracting the
background from our 8 events in the signal region, we determine a
branching fraction using a detection efficiency for the single
muon of 69.9\% and 28574$\pm$207 $D^{\mp}$ tags. We find
${\cal{B}}(D^+\to\mu^+\nu_{\mu})=(3.5\pm 1.4 \pm 0.6)\times
10^{-4}$ and the decay constant is $f_{D^+}=(198\pm 41\pm 17)~{\rm
MeV}$. \indent I gratefully acknowledge my advisor Professor S.
Stone for all his guidance and help.


\begin{thebibliography}{0}









\bibitem{Formula1}
   J. L. Rosner, in {\bf Particles and Fields 3}, Proceed. of the 1988 Banff
Summer Inst., Banff, Alberta, Canada, ed. by A. N. Kamal and F. C.
Khanna, World Scientific, Singapore, 1989, 395.


\bibitem{PDG}
S. Eidelman \etal ~(PDG), Phys. Lett. B {\bf 592}, 1 (2004).

\bibitem{detector}CLEO Collaboration,
                  CLNS-01/1742 and references therein.
\bibitem{ICHEP:Nabil}
D. Besson \etal, ''Measurement of ${\cal{B}}(D^+\to\mu^+\nu)$ and
the Pseudoscalar Decay Constant $f_{D^+}$", [hep-ex/[0408071].


















\end{thebibliography}
\end{document}